\documentclass[12pt]{JHEP3}

\usepackage{amsmath,epsfig}
\usepackage{amssymb,amsfonts}
\usepackage{latexsym}
\relax
\renewcommand{\theequation}{\arabic{section}.\arabic{equation}}
\def\be{\begin{equation}}
\def\ee{\end{equation}}
\def\bea{\begin{eqnarray}}
\def\eea{\end{eqnarray}}

\newcommand\fverb{\setbox\pippobox=\hbox\bgroup\verb}
\newcommand\fverbdo{\egroup\medskip\noindent%
                        \fbox{\unhbox\pippobox}\ }
\newcommand\fverbit{\egroup\item[\fbox{\unhbox\pippobox}]}

\newcommand{\bear}{\begin{eqnarray}}

\newcommand{\eear}{\end{eqnarray}}

\newbox\pippobox

\def\6{\partial}

\def\a{\alpha}

\def\pa{\partial}

\def\e{\epsilon}

\def\s{\sigma}

\def\sp{\;\;\;,\;\;\;}

\def\sq
\def\a{\alpha}

\def\l{\lambda}

\def\hri#1#2{\href{http://arxiv.org/abs/#1}{[ArXiv:#1]#2}}
\def\hre#1#2{\href{http://arxiv.org/abs/#1/#2}{[ArXiv:#1/#2]}}

\def\na{\nabla}

\def\e{\epsilon}

\title{Spherically symmetric solutions in modified Ho\v rava-Lifshitz gravity. }
\author{{\large Elias Kiritsis\footnote{On leave
of absence from APC, Universit\'e Paris 7, (UMR du CNRS 7164)}}
~\\
~\\
Crete Center for Theoretical Physics,
Department of Physics,\\ University of Crete,
71003 Heraklion, Greece}

\preprint{CCTP-2009-20~~~~~~~~~~~~~~\\arXiv:0911.3164 [hep-th]}

\abstract{We find spherically symmetric solutions in the modified Ho\v rava-Lifshitz gravity
proposed recently by Blas, Pujolas and Sibiryakov. The non-linear equations of the two derivative action turn out to be similar to those
stemming from the
four-derivative action explored recently.
We analyze the solutions and derive constraints on the relevant new coupling constant.
We also analyze the case where the cosmological constant is non-zero.
 We derive the large distance expansion of solutions and show that the power of the standard Newton's law is modified in the
 presence of a cosmological constant.}

{
\begin{document}

\section{Introduction and results}

Ho\v rava provided recently with an alternative view on the quantization of gravitational theories, \cite{hor1}. By abandoning
 Lorentz invariance he was able to allow for a power counting renormalizable class of theories.
Several aspects of such theories have been analyzed since, and such progress was reviewed in \cite{kk} where we refer the reader
for references and more details.
A generic problem of the theory proposed by Ho\v rava is the  presence of unstable modes as well as strong coupling directions, as it usually happens
 in non-diffeomorphism invariant theories. It was proposed in \cite{blas2} that the addition of spatial derivatives of the lapse function can
 ameliorate the behavior of the theory, and indeed it was shown that the linearized theory around
 flat space has proper behavior for a range of the extra couplings.

The purpose of the present paper is to study non-linear aspects of the  theories proposed in \cite{blas2}
by studying its  spherically symmetric solutions.
It turns out that the resulting equations for arbitrary couplings are much more complicated than those studied (and solved) recently in \cite{kk}.
To proceed we will simplify the theory so that it is still tractable while it contains the new ingredients proposed in \cite{blas2}.

For this we will choose the action to be
\be S=\int
dtd^3x\sqrt{g}N\!\!\left[\alpha (K_{ij}K^{ij}\!-\!\l K^2)+\beta C_{ij}C^{ij}\!+\xi R+a_1 (a_ia^i)\right]\sp a_i\equiv {\partial_iN\over N}
\label{q1}
\ee
and we will look for spherically symmetric solutions with zero shift of the form
\be
ds^2=-N(r)^2dt^2+{dr^2\over f(r)}+r^2d\Omega^2
\label{q2}\ee
Defining the dimensionless variable $b={4\xi\over a_1}$, the linearized stability of the theory constraints $0<{a_1\over \xi}<2$ or $b>2$.

The nonlinear equations are solved and we find the following types of solutions:
\begin{itemize}

\item In the stable range,  $b>2$, we find an asymptotically flat solution with positive ``mass",
 which extends down to a regular horizon\footnote{This will be a horizon only for regular relativistic matter,
 see the discussion in \cite{kk}.} (defined as the solution
$r_*$ to $f(r_*)=0$).
The asymptotic expansion of this solution is
\be
f=1-{2GM\over r}-{(2GM)^2\over 2b r^2}-{(2GM)^3\over 4br^3}+{\cal O}(r^{-4})
\label{pp1n}\ee
\be
N^2=1-{2GM\over r}+{2(GM)^3\over 3br^3}+{\cal O}(r^{-4})
\label{pp2n}\ee
\be
2GM={2\over 1+\sqrt{b\over b-2}}\left({\sqrt{b}-\sqrt{b-2}\over
\sqrt{b}+\sqrt{b-2}}\right)^{1\over b-2+\sqrt{(b-2)b}}~r_*
\label{pp3n}\ee
This expansion  defines the parameter we call ``mass".
There is an infinite series of subleading terms controlled by the mass and the parameter $b$.
In particular, the subleading behavior in $N^2$ has the same power  as  the one induced in RS compactifications.
There is however an important qualitative difference in the sense that in RS the coefficient is linear in the mass while
here it is cubic in the mass. Therefore,  the effects of the new interaction are more prominent for very massive objects.
As we vary the parameter $2<b<\infty$, the ratio $r_*/2GM$ in (\ref{pp3n}) varies from 1 (at $b=\infty$ where the interactions are not present)
 to $\infty$ at $b=2$.

There is another class of solutions with non-trivial asymptotic behavior
\be
f\sim r^{b-2+\sqrt{b(b-2)}}\sp N\sim r^{-{1\over 2}\sqrt{b\over b-2}}
\ee
as $r\to \infty$. For such solutions, the four-dimensional scalar curvature is singular as $r\to \infty$.

\item In the borderline case $b=2$, we find solutions with $f(r)$ arbitrary and
\be
{rN'\over N}=-1+{\e\over \sqrt{f}}\sp \e^2=1
\label{pp50}\ee
This is in tune with the degeneracy of the spatial derivative action of the scalar mode in this case, \cite{blas2}.

\item In the unstable range, $0<b<2$, we also find an asymptotically flat solution with positive ``mass", which extends down to a regular horizon.
Its large distance expansion is similar to (\ref{pp1n}), (\ref{pp2n}).
There are no exotic asymptotics in this case.

\item We also analyze the case where a cosmological constant is present. Although we cannot solve the
 non-linear equations in that case, we can solve them for the large
distance expansion. We classify the large distance asymptotics in that case.
 We find the regular solutions have an expansion that is
given by
\be
f=2-{2\xi\over 2\xi-a_1}+{(\xi-a_1)^2\over \xi(3\xi-2a_1)(2\xi-a_1)}\s r^2+{C\over r^{a-2\over a+2}}+\cdots\sp a=2-4{\xi\over a_1}
   \ee
where $f$ is the standard blackness function and $C$ is a constant of integration.
We observe that the $1/r$ tail of the solutions without a cosmological constant is now changed to $1/r^{a-2\over a+2}$.
This implies that the power of Newton's law is modified in the presence of a cosmological constant.

\end{itemize}

Such solutions can be understood as long range tails of matter distributions (stars).
The presence of a physical singularity in such solutions needs to be further understood. As argued in \cite{kk} the curvature invariants are not enough
to ascertain singular physics. This has been also shown recently \cite{muko} by analyzing spherically symmetric solutions in the projectable theory.

A theory with action (\ref{q1}) differs from General Relativity already at the two-derivative level. Although this difference is at the
level of gravitational (and not matter interactions) we expect constraints on the coefficient $a_1$ to be mild.
We can estimate that the relative change to the perihelion precession for Mercury is about $-{5\over b}10^{-8}$.
 Since $b>2$ in the stable range, such a constraint is
comfortably irrelevant due to the non-relativistic suppression. The most stringed constraints on $b$ will be set by
binary pulsar data and are expected to be in the $10^{-3}$ range.

\section{The Ho\v rava-Lifshitz gravity theory and its generalizations}

We review here the gravitational theories of the Ho\v rava-Lifshitz type with general couplings.

The dynamical variables are $N,N_i,g_{ij}$, with (mass) scaling dimension\footnote{Note that this is the canonical scaling
guaranteeing power-counting renormalizability once the dimensionfull constant in front of $dt^2$ is set to one.}
0, except $N_i$ that has scaling dimension 2. This is similar to
the ADM decomposition of the metric in standard general relativity,
where the metric is written as
\be ds^2=-N^2
~dt^2+g_{ij}(dx^i+N^idt)(dx^j+N^jdt)\sp N_i=g_{ij}N^j. \label{q3}
\ee

The most general power-counting renormalizable action is
\be S=S_{kin}+S_{3}+S_{rel}+S_{new}
\label{q10}\ee
\be
S_{kin}=\alpha \int
dtd^3x\sqrt{g}N\!\!\left[(K_{ij}K^{ij}\!-\!\l K^2)\right]
\label{q4}\ee
\be
S_{rel}=\int
dtd^3x\sqrt{g}N\left[\gamma
{\cal E}^{ijk}R_{il}\na_j{R^l}_{k} \!+\!\zeta R_{ij}R^{ij}\!+\!\eta
R^2\!+\!\xi R\!+\!\sigma\!\right],
\label{q5}\ee
\be
S_3=\int dtd^3x\sqrt{g}N\left[ \beta C_{ij}C^{ij}+\beta_1 R\square
R+\beta_2R^3+\beta_3RR_{ij}R^{ij}+\beta_4
R_{ij}R^{ik}{R^{j}}_k\right]
\label{q6}\ee
\be
S_{new}=\int dtd^3x\sqrt{g}N\left[a_1 (a_ia^i)+a_2 (a_ia^i)^2+a_3R^{ij}a_ia_j+a_4R\nabla_i a^i+
\right.
\label{q7}\ee
$$
\left.+a_5\nabla_ia_j\nabla^ia^j+
a_6\nabla^i a_i (a_ja^j)+\cdots
\right]
$$
The extrinsic curvature is defined as
\be
K_{ij}={1\over 2N}(\dot
g_{ij}-\nabla_{i}N_j-\nabla_jN_i), \sp
 K=g^{ij}K_{ij}\sp
K^{ij}=g^{ik}g^{jl}K_{kl}
\label{q8}\ee
 with covariant
derivatives defined with respect to the spatial metric $g_{ij}$.
$ {\cal
E}^{ijk}={\e^{ijk}\over \sqrt{g}}$ the standard generally covariant
antisymmetric tensor. $\e^{123}$ is defined to be 1, and other
components are obtained by antisymmetry. Indices are raised and
lowered with the metric $g_{ij}$. Therefore, ${\cal E}^{ijk}=(\pm 1)
/\sqrt{g}$. The Cotton tensor in (\ref{q6}) is given by
 \be C^{ij}={\cal E}^{ikl}\na_k\left({R^j}_{l}-{1\over 4}R{\delta^{j}}_{l}\right)
\label{q9}\ee

In (\ref{q7}) we have defined
\be
 a_i\equiv {\partial_iN\over N}
 \label{q11}
 \ee
and the ellipsis in (\ref{q7}) refers to dimension six terms involving $a_i$ as well as curvatures.

The action in (\ref{q10}) is invariant under a restricted class of
diffeomorphisms
\be t'=\hat h(t)\sp x'^{\,i}=\hat h^i(t,\vec x). \label{16}\ee
The transformation of the metric under infinitesimal diffeomorphisms ($\hat h\simeq 1+k$, $\hat h^i\simeq x^i+\e^i$)
is \be \delta
g_{ij}=\pa_i\e^kg_{jk}+\pa_j\e^kg_{ik}+\e^k\pa_kg_{ij}+k\dot g_{ij}
\ee \be \delta N_i=\pa_i\e^jN_j+\pa_j\e^jN_i+\dot \e^j g_{ij}+\dot k
N_i+k\dot N_i\sp \delta N =\e^j\pa_j N+\dot k N+k{\dot N}. \ee

The equations of motion stemming from the action (\ref{q10}) are formidable, even for simple ansatze like the spherically symmetric ansatz
that will be considered here\footnote{The only ansatz that is still  tractable is the homogeneous and isotropic cosmological ansatz whose solutions
are the same as those discussed in \cite{kkco}.}.
We will therefore simplify the theory, keeping however some of its main features. We will concretely study the action
\be
S=\int
dtd^3x\sqrt{g}N\!\!\left[\alpha (K_{ij}K^{ij}\!-\!\l K^2)+\beta C_{ij}C^{ij}\!+\xi R+a_1 (a_ia^i)\right]
\label{qq1}
\ee
which contains the kinetic terms, the leading conformally invariant dimension 6 terms, the Einstein term necessary for the IR limit
and the leading relevant term of the new kind advocated in \cite{blas2}.
It should be mentioned that the last term breaks Lorentz invariance in a important way and therefore we expect its coefficient to be constrained.
As it is in the gravitational sector the constraints are expected to be mild.

We will search for static, spherically symmetric solutions with zero shift of the form
\be
ds^2=-N(r)^2dt^2+{dr^2\over f(r)}+r^2d\Omega^2
\label{qq2}\ee
For such solutions the kinetic and Cotton terms do not contribute to the equations of motion.

\section{The equations of motion}

We insert the ansatz (\ref{qq2}) into the equations of motion.
The equation stemming from the variation of $N$  becomes
\be
{1\over r^2}\left[2\xi\left(1-f-rf'\right)-a_1\left(4f+rf'\right){rN'\over N}+a_1f{r^2N'^2\over N^2}-2a_1f{r^2N''\over N}\right]=0
\label{p8}\ee
The equation stemming from varying $g_{11}$ is
\be
r^2\left[4\xi(f-1)+8\xi f{rN'\over N}+2a_1f{r^2N'^2\over N^2}\right]=0
\label{p9}\ee
 while the one stemming from $g_{22}$ variation is
 \be
 r^2\left[-2\xi rf'-2\xi \left(rf'+2f\right){rN'\over N}-4\xi f{r^2N''\over N}+2a_1 f{r^2N'^2\over N^2}\right]=0
 \label{p10}\ee
 All other equations are trivially satisfied.

 It is convenient at this point to define the following parameters
\be
w=a_1-2\xi\sp  a={2w\over a_1}\sp b={4\xi\over a_1}\sp a+b=2
 \ee
For a positive Planck scale we must have $\xi>0$. Stability of linear fluctuations around flat space implies $0<{a_1\over \xi}<2$
that translates into $b>2$ and consequently $a<0$. The general relativity equations are recovered in the limit $a_1\to 0$ or $b\to \infty$.

We now proceed to solve (\ref{p8}) for  $N''$ and substitute in (\ref{p10}) to obtain the equation
\be
{a_1}f(\xi-a_1){r^2N'^2\over N^2}-2a_1\xi {f}{rN'\over N}+(a_1-2\xi)\xi{rf'}+{2\xi^2}(1-f)=0
\label{p11}\ee
We now combine this with (\ref{p9}) to eliminate $N$ and obtain a differential equation purely for $f$.
This is valid as long as $a\not= 0$. We will return to the $a=0$ case later.
The equation for $f$ reads
\be
r^2f'^2+2a rff'-4rf'+2af^2-2(2+a)f+4=0
\label{p13}\ee

The strategy is to first solve (\ref{p13}) and determine $f$, and then solve (\ref{p9}), determine $N'/N$ algebraically in terms of $f$
and then integrate once to obtain $N(r)$. As there are three equations for two unknown functions, the solutions of the procedure above must be finally substituted to the
original equations in order to verify whether there are further constraints.

\section{Large distance expansions}

Before proceeding with the general solution of (\ref{p13}) we will investigate the asymptotic expansion of asymptotically flat solutions when they exist.
Assuming that at large $r$ $f$ asymptotes to a constant (that we normalize to one),
 we can show that the equation (\ref{p13}) admits the following large distance expansion
\be
f=1-{2GM\over r}-{(2GM)^2\over 2b r^2}-{(2GM)^3\over 4br^3}+{\cal O}(r^{-4})
\label{pp1}\ee
\be
N^2=1-{2GM\over r}+{2(GM)^3\over 3br^3}+{\cal O}(r^{-4})
\label{pp2}\ee
where $GM$ is an integration constant that we interpreted as the Newton constant times the mass.
 Note that the corrections to the Newton's law are controlled both by the mass and the parameter $b$ that is controlled by the non-standard term
 in the action (\ref{qq1}).

\section{Finding Solutions}
To proceed further we define
\be
f=1+{b\over a}+g(r)
\label{p14}\ee
so that (\ref{p13}) becomes
\be
r^2g'^2+2a (rgg'+g^2)+2b g=0
\label{p15}\ee
We further define
\be
g=r^{-a}h\sp r=e^u
\label{p16}\ee
to rewrite (\ref{p15}) as
\be
\dot h^2+abh^2+2be^{au}h=0
\label{p18}\ee
This equation is similar but not identical to the one solved in \cite{kk}. It turns out that a similar technique applies here, namely we
 define
\be
W={h\over \dot h}
\label{p19}\ee
so that
\be
h=-{2b~e^{au}W^2\over 1+ab W^2}\sp g=-{2b~W^2\over 1+ab W^2}
\label{p21}\ee
where we have used (\ref{p18}).

The N function can be obtained from (\ref{p9}) that can be rewritten as
\be
\left({rN'\over N}\right)^2+b{rN'\over N}+{b\over 2}{f-1\over f}=0
\ee
with solution
\be
{rN'\over N}=-{b\over 2}\left[1+\e a W\right]\sp \e^2=1
\label{p21a}\ee
Substituting (\ref{p14}), (\ref{p21}), (\ref{p21a}) into the three equations (\ref{p8})-(\ref{p10})
we obtain that that they are solved iff $W$ satisfies
\be
2\e\dot W=-\left(1+a\e W\right)(1+abW^2)
\label{p20}\ee
We note that the equation for $\e=1$ above is obtained from the equation with $\e=-1$ by $W\to -W$.
As such a sign change does not change $f$, we restrict ourselves for now on, without loss of generality, to the $\e=-1$ case.

To proceed further and integrate (\ref{p20}) we have to distinguish three cases

\section{Study of the solutions for  $a<0$}

\begin{figure}[ht]
\begin{center}
\includegraphics[width=14cm]{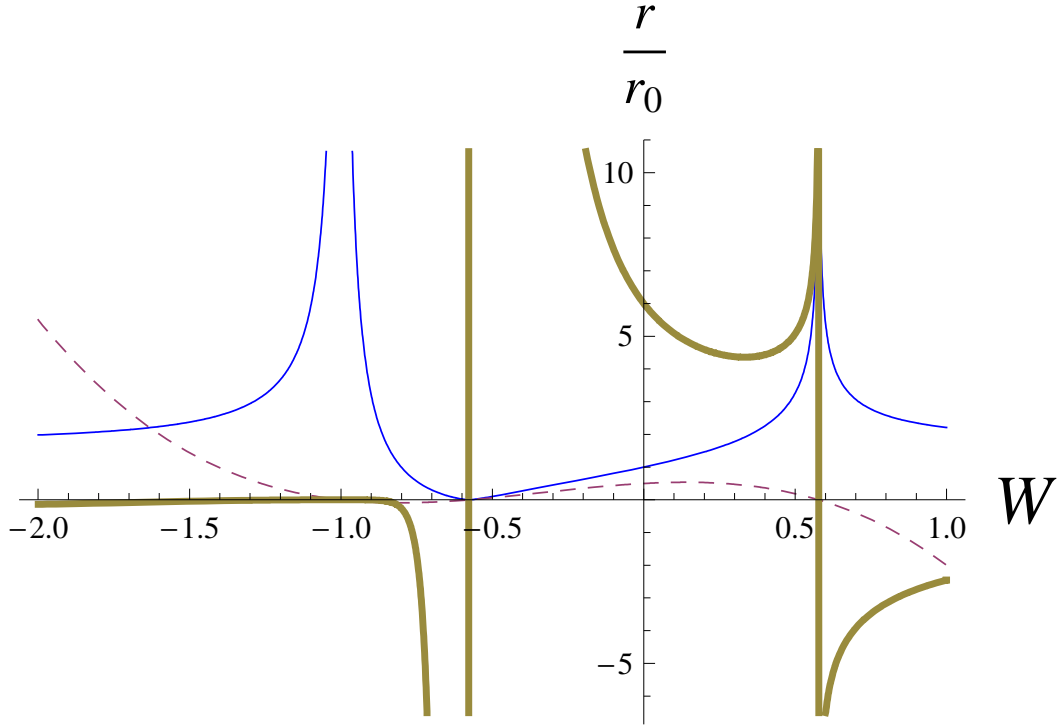}
\end{center}
\caption{\em Solid lines: $r/r_0$ (vertical axis) is plotted against $W$ (horizontal  axis), for $a=-1,b=3$.
There are four branches that reach asymptotic infinity at $W=-{1\over |a|}$ and $W={1\over \sqrt{|a|b}}$.
$r$ reaches zero when $W=-{1\over \sqrt{|a|b}}$.
Dashed line:  $r/r_0$ (vertical axis) is plotted against $\dot W$. Thick line: The four-dimensional scalar curvature (in units of $r_0^{-2}$). }
\label{fig0}
\end{figure}

\begin{figure}[ht]
\begin{center}
\includegraphics[width=9.5cm]{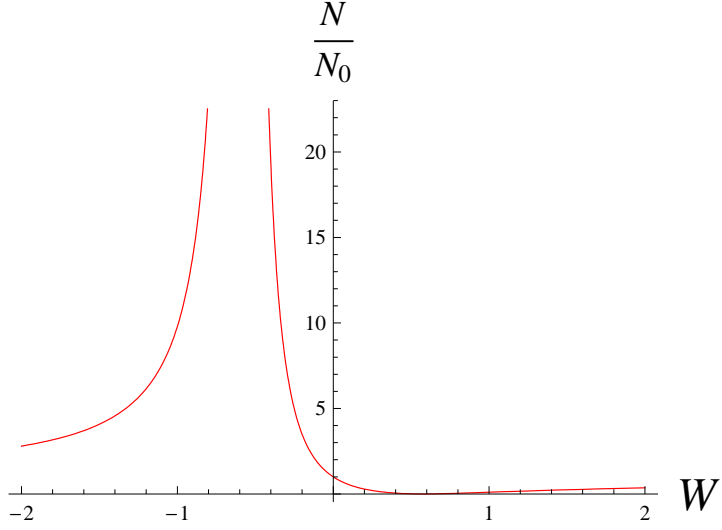}
\end{center}
\caption{\em $N/N_0$ (vertical axis) is plotted against $W$ (horizontal  axis), for $a=-1,b=3$.}
\label{fig00}
\end{figure}

This is the regime where the linearized theory is well behaved.
In this case the solution is
\be
{\Big |1+\sqrt{|a|b}~W\Big |^{1+{1\over \sqrt{|a|b}+|a|}}
\over \Big |1+|a|~W\Big |\Big |1-\sqrt{|a|b}~W\Big |^{{1\over \sqrt{|a|b}+|a|}}}={r\over r_0}
\label{p26}\ee
while
\be
{N\over N_0}=\Big |{1-\sqrt{|a|b}W\over 1+\sqrt{|a|b}W}\Big |^{{1\over 2}\sqrt{b\over |a|}}
\label{p25}\ee

In figue \ref{fig0} the solution $W(r)$ is plotted in inverted form: $r/r_0$ as a function of $W$.
$N/N_0$ is plotted as a function of $W$ in figure \ref{fig00}.

Rewriting the solution in terms of $g(r)$ we must distinguish two cases

$\bullet$ $W=\sqrt{|a|g\over |a|g-2}$ in which case we obtain
\be
{\Big |1+\sqrt{|a|g\over |a|g-2}\Big |^{1+{1\over \sqrt{|a|b}+|a|}}\over
\Big |1+\sqrt{|a|\over b}\sqrt{|a|g\over |a|g-2}\Big |\Big |1-\sqrt{|a|g\over |a|g-2}\Big |^{{1\over \sqrt{|a|b}+|a|}}}={r\over r_0}
\label{p241}\ee
\be
{N\over N_0}=\Big |{1-\sqrt{|a|g\over |a|g-2}\over 1+\sqrt{|a|g\over |a|g-2}}\Big |^{{1\over 2}\sqrt{b\over |a|}}
\label{p251}\ee

\begin{figure}[ht]
\begin{center}
\includegraphics[width=7cm]{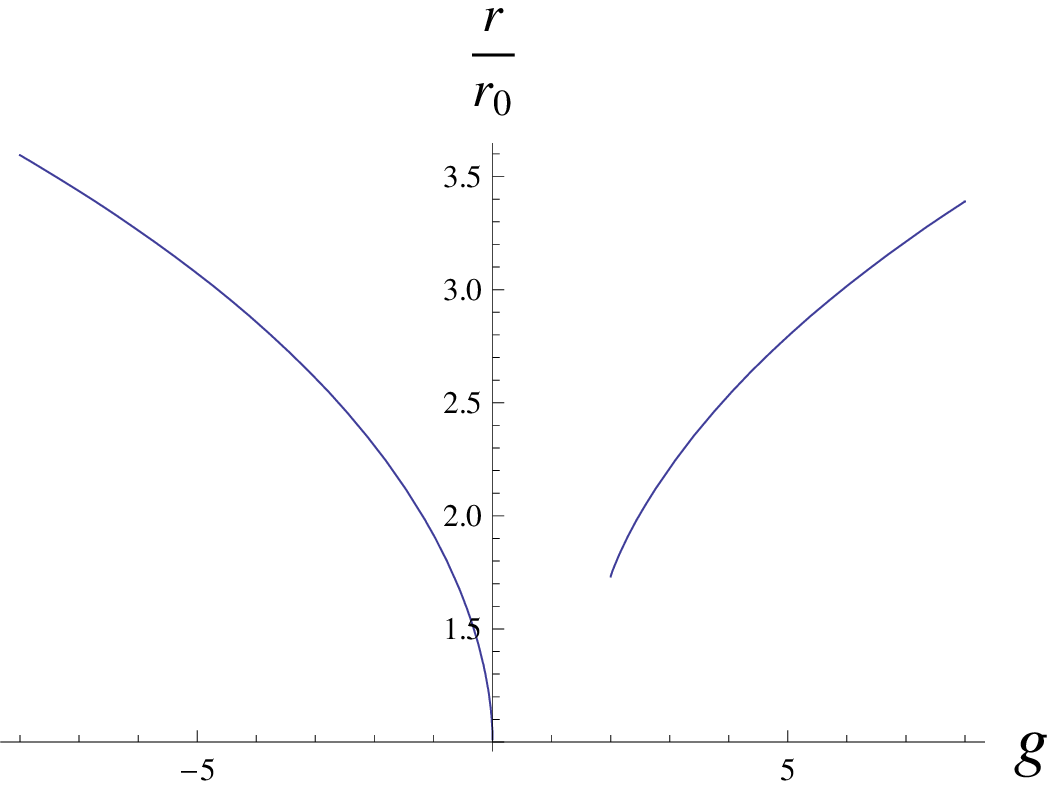}~~~~\includegraphics[width=7cm]{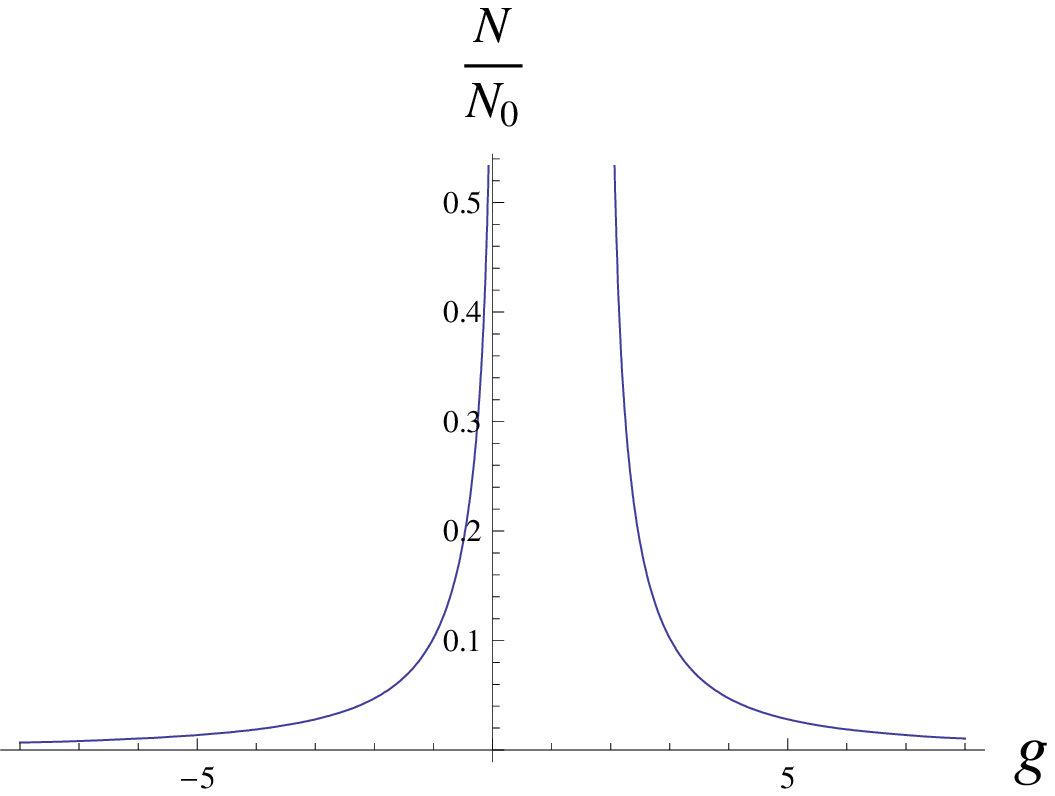}
\end{center}
\caption{\em Left figure: $r/r_0$ (vertical axis) is plotted against $g$ (horizontal  axis), for $a=-1,b=3$ when $W>0$.
The left branch has $g<0$ while the right branch has $g>{2\over |a|}$. Asymptotic infinity $r\to\infty$ is at $g\to \pm \infty$.
In the left  branch $r_0<r<\infty$ while in the right branch $\sqrt{b\over |a|}r_0<r<\infty$.
The right branch corresponds to ${1\over |a|b}<W<\infty$, while the left branch to  $0<W<{1\over |a|b}$.
Right figure:$N/N_0$ (vertical axis) is plotted against $g$ (horizontal axis), for $a=-1,b=3$.}
\label{fig1}
\end{figure}

The two branches are described in figure \ref{fig1}.
Both have a region at infinity and as we will now see they have non-standard large distance asymptotics.

For $r\to\infty$ we must have $|ag|\to \infty$. Expanding around this limit we obtain:
\be
|a|g\simeq \pm\left({\left(1+\sqrt{|a|\over b}\right)r\over 2^{1+{1\over \sqrt{|a|b}+|a|}}r_0}\right)^{\sqrt{|a|b}+|a|}+\cdots\sp
{N\over N_0}\simeq \left({2^{{1\over \sqrt{|a|b}+|a|}}r_0\over \left(1+\sqrt{|a|\over b}\right)r}\right)^{{1\over 2}\sqrt{b\over |a|}}+\cdots
\label{p33}\ee

The metric is therefore not asymptotically flat.
Moreover as seen from figure \ref{fig0} there the four-dimensional scalar curvature diverges in this limit.

\begin{figure}[ht]
\begin{center}
\includegraphics[width=7cm]{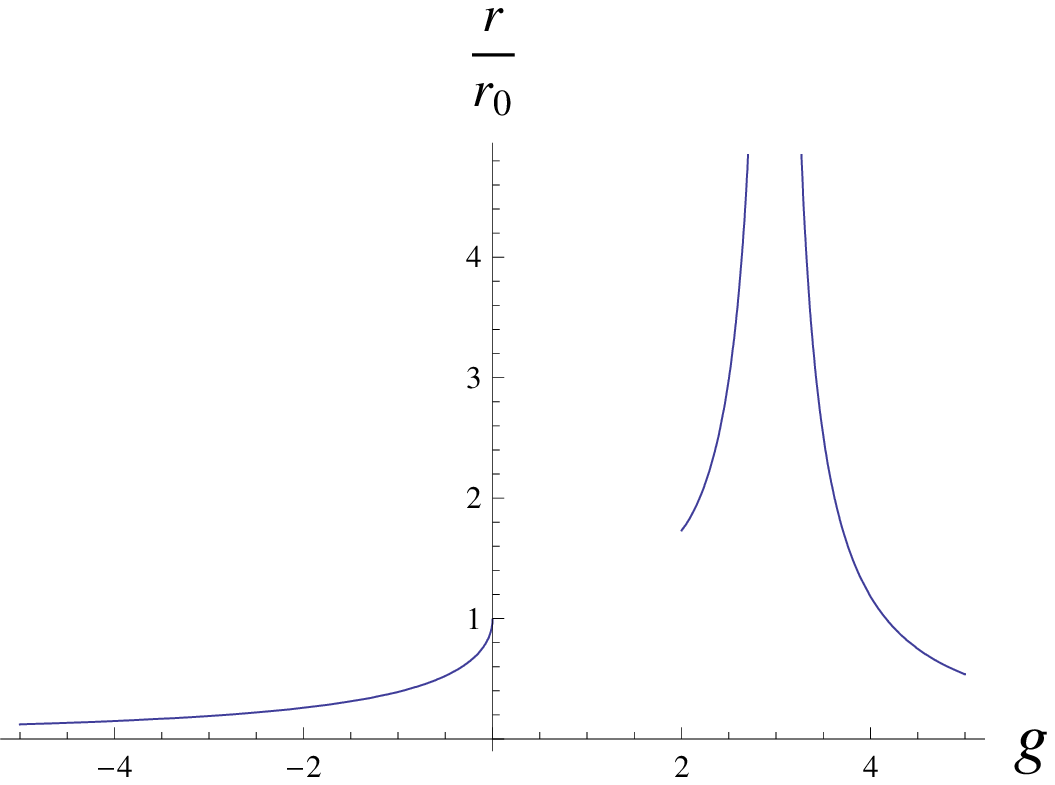}~~~\includegraphics[width=7cm]{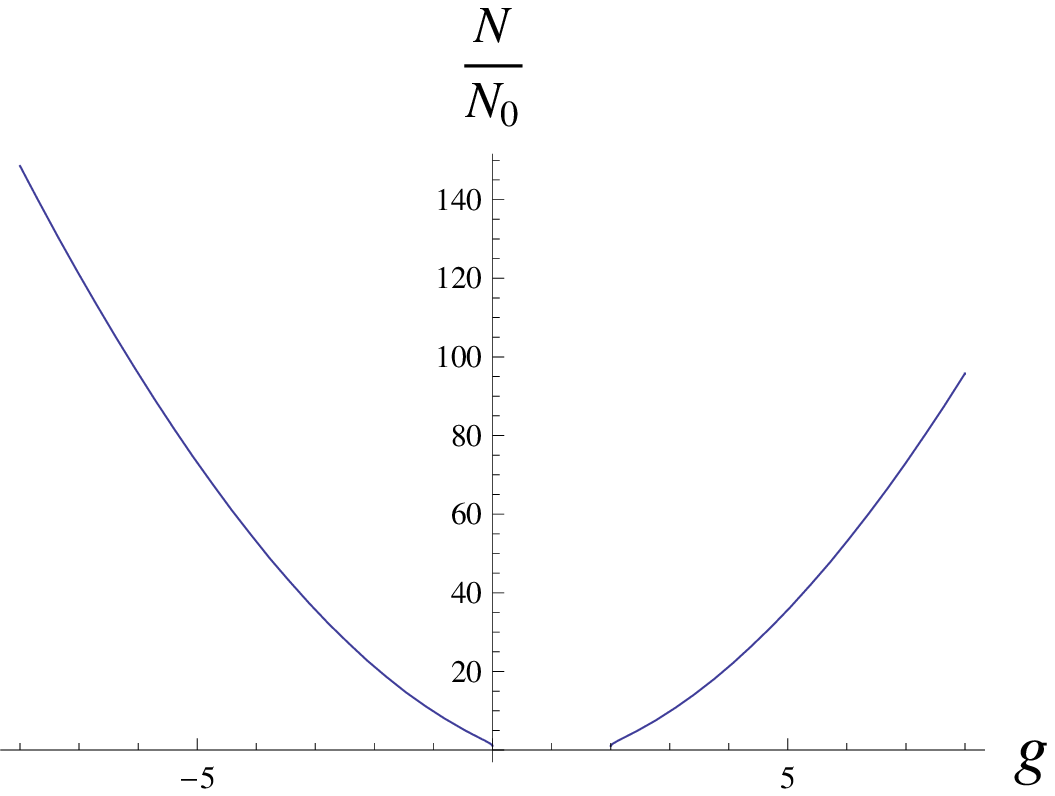}
\end{center}
\caption{\em Left figure: $r/r_0$ (vertical axis) is plotted against $g$ (horizontal  axis), for $a=-1,b=3$ when $W<0$.
The left branch has $g<0$, the center  branch has ${b\over |a|}>g>{2\over |a|}$ while the right branch has $g>{b\over |a|}$.
 Asymptotic infinity $r\to\infty$ is at $g={b\over |a|}$.
In the left  branch $0<r<r_0$, in the  center branch $\sqrt{b\over |a|}r_0<r<\infty$ while in the right branch $0<r<\infty$.
The left branch corresponds to $-{1\over \sqrt{|a|b}}<W<0$, the right branch to  $-{1\over |a|}<W<-{1\over \sqrt{|a|b}}$
and the  center branch to $-{1\over |a|}>W>-\infty$.
Right figure:$N/N_0$ (vertical axis) is plotted against $g$ (horizontal axis), for $a=-1,b=3$, $W<0$.}
\label{fig2}
\end{figure}

$\bullet$ $W=-\sqrt{|a|g\over |a|g-2}$ in which case we obtain
\be
{\Big |1-\sqrt{|a|g\over |a|g-2}\Big |^{1+{1\over \sqrt{|a|b}+|a|}}\over
\Big |1-\sqrt{|a|\over b}\sqrt{|a|g\over |a|g-2}\Big |\Big |1+\sqrt{|a|g\over |a|g-2}\Big |^{{1\over \sqrt{|a|b}+|a|}}}={r\over r_0}
\label{p2411}\ee
\be
{N\over N_0}=\Big |{1+\sqrt{|a|g\over |a|g-2}\over 1-\sqrt{|a|g\over |a|g-2}}\Big |^{{1\over 2}\sqrt{b\over |a|}}
\label{p2511}\ee

As seen in figure \ref{fig2} this solution contains three branches corresponding to physically different solutions.

\begin{itemize}

\item
The right branch has $g>{b\over |a|}$ and $r$ varies from 0 to $\infty$.
In particular, $g(r=0)=\infty$ and $g(r=\infty)={b\over |a|}$ that corresponds to $f=1$.
It corresponds to $-{1\over |a|}<W<-{1\over \sqrt{|a|b}}$. As $g>{b\over |a|}$we will soon see that this branch corresponds to negative masses.

\item The center branch has  ${2\over |a|}<g<{b\over |a|}$ and $r$ varies from ${b\over |a|}r_0$ to $\infty$.
 In particular, $g\left(r={b\over |a|}r_0\right)={2\over |a|}$ that corresponds to
 $f\left(r={b\over |a|}r_0\right)=0$ and $g(r=\infty)={b\over |a|}$ that corresponds to $f=1$.
It corresponds to $-\infty<W<-{1\over |a|}$. As $g<{b\over |a|}$ this branch corresponds to positive masses and is therefore
 interesting. Moreover, as can be seen from figure \ref{fig0} for this solution the scalar curvature is everywhere regular.

\item The left branch has  $g<0$ and $r$ varies from 0 to $r_0$.
 In particular, $g\left(0\right)=\infty$ and $g(r=0)={0}$.
It corresponds to $-{1\over \sqrt{|a|b}}<W<0$. This branch has $g<0$ but no asymptotic region.
It could however be patched together with the left branch on figure \ref{fig1} to make a single geometry with $0<r<\infty$.

\end{itemize}

The large distance expansion of the center branch is
\be
g(r)={b\over |a|}-{2b\over |a|+\sqrt{|a|b}}\left({\sqrt{b}-\sqrt{|a|}\over
\sqrt{b}+\sqrt{|a|}}\right)^{1\over |a|+\sqrt{|a|b}}{r_0\over r}+{\cal O}(r^{-2})
\ee
\be
N^2=1-{2b\over |a|+\sqrt{|a|b}}\left({\sqrt{b}-\sqrt{|a|}\over
\sqrt{b}+\sqrt{|a|}}\right)^{1\over |a|+\sqrt{|a|b}}{r_0\over r}+{\cal O}(r^{-2})
\ee
where we chose $N_0=\left({\sqrt{b}-\sqrt{|a|}\over
\sqrt{b}+\sqrt{|a|}}\right)^{{1\over 2}\sqrt{b\over |a|}}$.
This is full agreement with (\ref{pp1}) and (\ref{pp2}).

\section{Study of the solutions for  $a=0$}

In this case $w=0$ or $a_1=2\xi$. This situation is at the boundary of the range allowed in \cite{blas2}. Indeed in this case the term quadratic in the
spatial
derivatives of the scalar mode has a vanishing coefficient.

The three basic equations now read
\be
1-f-rf'-(4f+rf'){rN'\over N}+f{r^2N'^2\over N^2}-2f{r^2N''\over N}=0
\label{p47}\ee
\be
-1+f+2f{rN'\over N}+f{r^2N'^2\over N^2}=0
\label{p48}\ee
\be
1-f+rf'+rf'{rN'\over N}-3f{r^2N'^2\over N^2}+2f{r^2N''\over N}=0
\label{p49}\ee
The second equation is solved for
\be
{rN'\over N}=-1+{\e\over \sqrt{f}}\sp \e^2=1
\label{p50}\ee
The other two equations are identically satisfied if (\ref{p50}) is satisfied. Therefore the $a=0$ case is a degenerate case where
$f(r)$ is arbitrary while $N$ is given by (\ref{p50}).

This reflects the degeneracy in the quadratic  term for the scalar mode.

\section{Non-zero cosmological constant}

We will now proceed to understand solutions when the bare cosmological constant is non-zero.
The relevant action now is
\be
S=\int
dtd^3x\sqrt{g}N\!\!\left[\alpha (K_{ij}K^{ij}\!-\!\l K^2)+\beta C_{ij}C^{ij}\!+\xi R+a_1 (a_ia^i)+\s\right]
\label{qq22}
\ee
The independent equations are
\be
4w^3\xi^2f^2-4w^2\xi f\left[2\xi(w+\xi)+(2w+3\xi)\s r^2-wr\xi f'\right]
\label{p123}\ee
$$
+(w+2\xi)(2w\xi+2(w+\xi)\s r^2-wr\xi f')^2=0
$$
and
\be
4\xi(f-1)-2\sigma r^2+8\xi f{rN'\over N}+2(w+2\xi)f{r^2N'^2\over N^2}=0
\label{p124} \ee
Once (\ref{p123}) is solved for $f$ then $N$ can be determined from (\ref{p124}) as
\be
{rN'\over N}=-{b\over 2}\left[1-\sqrt{1-{2\over b}\left(
1-{1\over f}-{c\over 2b} r^2\right)}\right]
 \label{p69}\ee
where we defined
\be
 a={2w\over (w+2\xi)}\sp b={4\xi\over (w+2\xi)}\sp c={4\s \over (w+2\xi)}\sp a+b=2
\label{p44}\ee
As before, it suffices to consider  one of the signs in front of the square root. .

To solve (\ref{p123}) we define
\be
f=1+{b\over a}+{c\over ab} r^2+g(r)
\label{p41}\ee
so that equation (\ref{p123}) becomes
\be
r^2g'^2+2a(rg'+ g)g+(2b+c r^2) g=0
\label{p42}\ee
One obvious generic solution is $g=0$.
Note that  $\s$ can be completely rescaled out of this equation by redefining the radial variable as $r^2\to {r^2\over \s}$.
The only thing that remains in such a case is the sign of $\s$. We will assume it to be positive (AdS), although similar remarks apply in the
dS case.

As before this equation can be simplified by the substitution
\be
g=-{(2b+cr^2)W^2\over 1+abW^2}\sp
{rN'\over N}=-{b\over 2}\left[1-\sqrt{2a^2b^3W^2+cr^2(cr^2+b^2)\over b^2(cr^2+2b)}\right]
\label{p52}\ee
to become
\be
2rW'+(1+abW^2)\left(1-{(a+2)cr^2+2ab\over cr^2+2b}W\right)=0
\label{p53}\ee
There is also another branch that is obtained by $W\to -W$ that does not affect $g$ and can be therefore ignored.
Unfortunately, we have not managed to solve this equation exactly. It is however amenable to numerical integration
straightforwardly.

We will study however here its large distance asymptotics that
 will provide some interesting clues on the physics in the presence of a vacuum energy.
 This study is not trivial as the equation in question is non-linear
 and typically several long-distance asymptotics exist.

\subsection{The large distance asymptotics}

To find the large distance asymptotics, we will concentrate at distances $r^2\gg \left|{2b\over c}\right|$.
In this region we can neglect the $2b$ term in equation   (\ref{p42}) to obtain
\be
r^2g'^2+2a(rg'+ g)g+c r^2 g=0
\label{p51}\ee
This equation can be solved exactly by the transformation
\be
g=-{cr^2~Y^2\over 1+abY^2}
\label{p55}\ee
to obtain the equivalent equation
\be
2rW'+(1+abW^2)\left(1-(a+2)W\right)=0
\label{p54}\ee
which is the large $r$ limit of (\ref{p53}).
This can be solved exactly, and we will focus in the physical branch, $a<0$.
The general solution is
\begin{figure}[ht]
\begin{center}
\includegraphics[width=8cm]{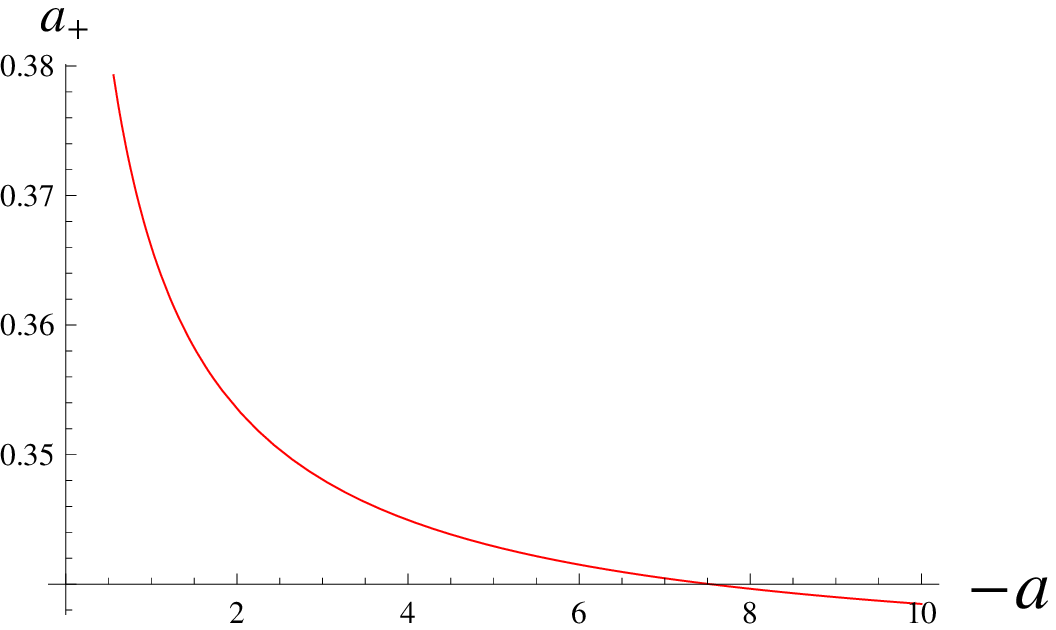}~~~\includegraphics[width=8cm]{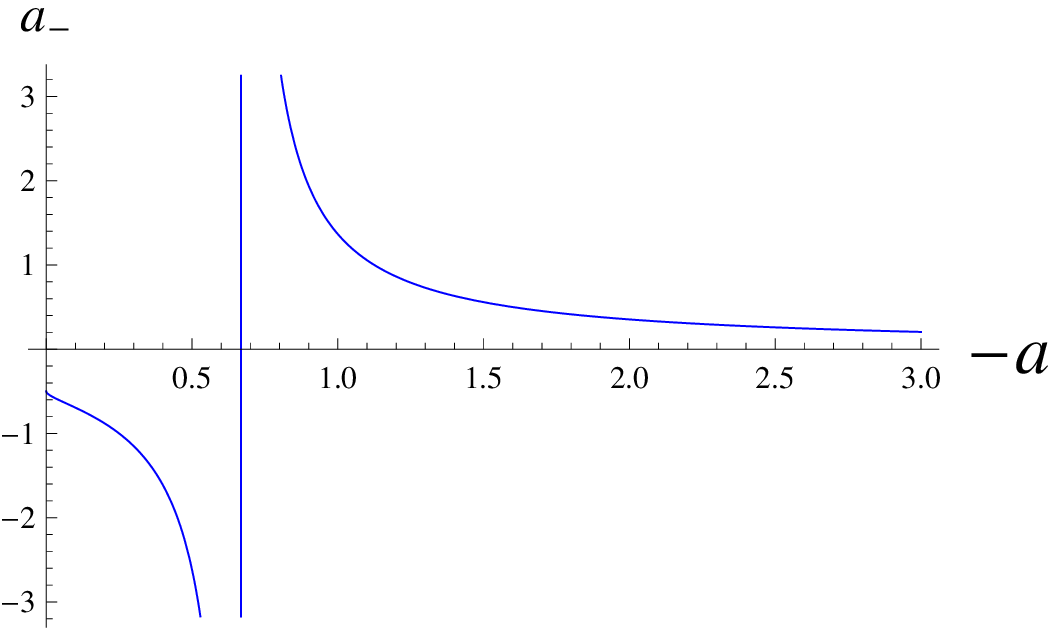}
\end{center}
\caption{Left: $a_+$ as a function of $|a|$. Right: $a_-$ as a function of $|a|$.}
\label{figc1}
\end{figure}

Integrating for $a<0$ we obtain
\be
{r\over r_0}
={\left|1+\sqrt{|a|b}Y\right|^{a_{+}}
\over \left|1-\sqrt{|a|b}Y\right|^{a_-}}~\left|{(|a|-2)Y+1}\right|^{(|a|-2)\over 2-3|a|}
\label{p56}\ee
with
\be
a_{\pm}={{\sqrt{|a|b}\pm (|a|-2)\over 6|a|-4}}
\label{p57}\ee

The values of $a_{\pm}$ for negative values of $a$ are plotted in figure  (\ref{figc1}).
We note that $a_{+}$ is always positive while $a_{-}$ is negative for $0<|a|<{2\over 3}$.

From (\ref{p56}) we observe that $r\to \infty$ on two distinct occasions.

\begin{enumerate}

\item $Y\to {1\over \sqrt{|a|b}}$ and $a_->0$.
This happens when $|a|>{2\over 3}$.
Expanding the solution we obtain
\be
Y={1\over \sqrt{|a|b}}+{\cal O}(r^{-{1\over a_-}})\sp g\sim r^{2+{1\over a_-}}
\label{p58}\ee

\item  $Y\to {1\over 2-|a|}$ and $|a|>2$ or $|a|<{2\over 3}$.
Here the asymptotics are
\be
Y\sim {1\over 2-|a|}+{\cal O}\left({1\over r^{3|a|-2\over |a|-2}}\right)\sp g\sim r^2
\label{p59}\ee

\end{enumerate}

Therefore
\begin{itemize}

\item for $0<|a|<{2\over 3}$, the only possible large-distance asymptotics is $g\sim r^2$ with subleading terms that are powers of $r^{3|a|-2\over |a|-2}$.

\item for ${2\over 3}<|a|<{2}$ the only large-distance asymptotics possible is $g\sim r^{2+{1\over a_-}}$.

\item finally for  $|a|>2$ both of the previous two asymptotics are possible.

\end{itemize}

The case $a=-{2\over 3}$ must be treated separately. For this value the solution (\ref{p56}) becomes
\be
{r\over r_0}=e^{-{9\over 4(4Y-3)}}\left|{4Y+3\over 4Y-3}\right|^{3\over 8}
\label{p60}\ee
To obtain $r\to\infty$  we must take $Y\to {3\over 4}^-$ and we obtain asymptotically
\be
Y\to {3\over 4}-{9\over 16\log r}+\cdots\sp
g\to -{3c\over 8}r^{2}\log r+\cdots
\label{p61}\ee

With the knowledge of the complete lists of asymptotics we can now proceed to establish the large distance expansions.

\subsection{The large distance expansions}

We start from solutions that asymptote to $r^2$ at large-distance.
The general large distance asymptotics is as follows.\footnote{We rescale r so that we set $c=1$.}
Solutions to (\ref{p42}) (with c=1) have the following large-distance structure
\be
g(r)=\sum_{n=0}^{\infty}~r^{2(1-n)+n{2-a\over 2+a}}~g_n(r)
\label{p62}\ee
where $g_0$ is given by
\be
g_0(r)=-{1\over 6a+4}+{1\over r^2}-{4(2+3a)\over (a+6)}{1\over r^4}
-{8 (a-2) (2 + 3 a)^2\over (6 + a) (10 + 3 a)} {1\over  r^6}+{\cal O}(r^{-6})
\label{p63}\ee
and in general
\be
g_n(r)=C_{n}\left[1+\sum_{m=1}^{\infty}{D_{n,m}\over r^{2m}}\right]
\label{p64}\ee
In particular, the first few terms are
\be
g_{1}(r)=C\left[1 - {2 (2 + 3 a)^2\over (2 + a)^2}{1\over r^2} - {
  2 (-2 + a) (2 + 3 a)^2 (4 + 28 a + 25 a^2 + 4 a^3)\over (2 + a)^4 (6 +
     a)}{1\over r^4}+{\cal O}(r^{-6})\right]
\label{p65}\ee
\be
g_2(r)=-{(2 + 3 a)^2 \over (2 + a)^2}C^2~\left[1 + {2 a(a-2) (2 + 3 a) (10 + 7 a)\over (2 + a)^2 (6 + 5 a)}{1\over
    r^2}+\right.
  \label{p66}  \ee
  $$
\left.  +{4 a(a-2)^2(2 + 3 a) (768 + 4160 a + 7288 a^2 + 5500 a^3 +
   1786 a^4 + 185 a^5)\over (2 + a)^4 (6 + a) (6 + 5 a) (10 + 7 a))}{1\over r^4}+{\cal O}(r^{-6})\right]
   $$
$C$ is an integration constant and $C_n\sim C^n$ in (\ref{p64}).
This expansion is well defined for generic irrational $a$. For rational values of $a$ resonance
 effects can happen and logarithms may appear in this expansion.

 From (\ref{p69}) we may compute the large distance expansion of the lapse. We find
 \be
 \log N^2={r^2\over 2}+(a-2)r+{a(a-2)\over 2}\log r+{(-2 + a)^2 a (-32 - 44 a + 4 a^2 + a^3)\over 16 (2 + a)^2 r^2}
\label{p70} \ee
$$
-{(a-2)^3a^2(3a+2)^2~C\over (a+2)^3(a+4)}r^{-3+{a-2\over a+2}}+\cdots
$$

For the blackness function $f$, we observe that this class of large distance asymptotics
have the standard $r^2$ behavior characteristic of gravity with a cosmological constant,
albeit the coefficient of the $r^2$ term is affected by the new term in the action.
Indeed in terms of the original couplings $\xi$, $a_1$ and $\s$, the blackness function $f$ from (\ref{p41}) has an expansion
\be
f=2-{2\xi\over 2\xi-a_1}+{(\xi-a_1)^2\over \xi(3\xi-2a_1)(2\xi-a_1)}\s r^2+\cdots
 \label{p67}\ee
that reduces to the standard asymptotics when $a_1=0$ (corresponding to $a\to -\infty$).

What is remarkable in  (\ref{p62}) is that no $1/r$ tail exists generically.
Indeed the leading term proportional to the integration constant behaves as $r^{2-a\over 2+a}$, and only becomes $1/r$
in the GR limit $a\to -\infty$.

The lapse on the other hand has completely different large distance asymptotics that are exponential to leading order, and where
the ``mass" contribution (proportional to $C$ in (\ref{p70}) is strongly subleading.

It is therefore expected that  the interplay between a small cosmological constant and the large distance tails of
spherically gravitating objects will be non-trivial, and unfortunately
is not easily accessible through a large distance expansion alone.
On the other hand, it is expected that there may be non-trivial constraints arising from this interplay
if the cosmological in the universe is assumed to be non-zero.

We now move to the other alternative large distance asymptotics described in the previous subsection.
In this case the solution can be written as
\be
g(r)=\sum_{n=1}^{\infty}r^{2+{n\over a_-}}~g_n(r)\sp g_n(r)=C_n\left(1+\sum_{m=1}^{\infty}{D_{n,m}\over r^{2m}}\right)
 \label{p68}\ee
$C_1$ is the integration constant while the other coefficients are non-linearly related to it.
The lapse has the same leading behavior as in (\ref{p70}).
Note that in the phenomenologically interesting region $a\to -\infty$, $1/a_-$ is large and positive.
Therefore this branch of solutions have a large power dependence which make it phenomenologically uninteresting.

\vspace{.7 in}
\addcontentsline{toc}{section}{Note Added}

\noindent {\bf Notes Added} \newline

\begin{itemize}

\item While this work was written, \cite{papa} has argued that the modified Ho\v rava-Lifshitz gravity proposed in \cite{blas2}
has also potential strong coupling regions, based on the two derivative approximation.
These regions are not visible in the spherically symmetric solutions studied here.
Subsequently in \cite{Blas3} it was argued that the breakdown of the two derivative theory is prevented by the higher derivative terms
rendering a theory without strong coupling problems.

\item It was also brought to our attention after the completion of this paper
that the equations solved for zero cosmological constant are identical with  those governing a special class of solutions
of a Einstein-Aether theory \cite{E3}, with
$c_1=c_3=0$, but with the aether vector restricted to be hypersurface orthogonal. This is the case with spherically symmetric solutions.
Indeed the solutions found here with $a<0$ match those found in \cite{E2}, and their linearized versions studied earlier in \cite{E1}.
It was also observed that the physical solution corresponding to the center branch of figure \ref{fig2} can be extended inside the horizon by the
right branch of the solution in figure \ref{fig1}. In that case the horizon at $f=0$ is a minimal sphere.
The relation with Einstein-Aether theory was also discussed in \cite{Blas3}.

\item After the appearance  of this paper, it was argued in \cite{Hen}, echoing an earlier claim in \cite{Li}
 that the second class constraints are very strong and in generic solutions they force the lapse to vanish. This does not happen in solutions with symmetry
 like those that are treated here, but this is certainly an important issue that deserves further attention.

\item After the completion of this work we became aware of reference \cite{poly} where spherically symmetric solutions
in standard Ho\v rava-Lifshitz gravity with non-zero shift were presented.

\end{itemize}

\vspace{.7 in}
\addcontentsline{toc}{section}{Acknowledgments}

\noindent {\bf Acknowledgements} \newline

We are grateful to C. Eling, T. Jacobson, G. Kofinas, A. Papazoglou, S. Sibiryakov, T. Sotiriou and H. B. Zang for
discussions and correspondence.

 This work was  partially supported by  a European Union grant FP7-REGPOT-2008-1-CreteHEP
 Cosmo-228644, and a CNRS PICS grant \# 4172.

\newpage
\appendix
\renewcommand{\theequation}{\thesection.\arabic{equation}}
\addcontentsline{toc}{section}{APPENDIX\label{app}}
\section*{APPENDIX}

\section{Study of the solutions for  $a>0$\label{a}}

We present in this appendix the solutions to the equations without a cosmological constant for $a>0$.

In this branch, as $a>0$ it implies that $b<2$, and therefore corresponds to the linearly unstable class,\cite{blas2}.

 The solution is
\be
{\sqrt{1+ab W^2}~\exp\left[\sqrt{b/a}\arctan(\sqrt{ab}W)\right]\over \Big |1-aW\Big |}
={r\over r_0}
\label{p27}\ee
while for N we obtain

\be
N=N_0\exp\left[-\sqrt{b\over a}{\arctan(\sqrt{ab}W)}\right]
\label{p29}\ee

\begin{figure}[ht]
\begin{center}
\includegraphics[width=7cm]{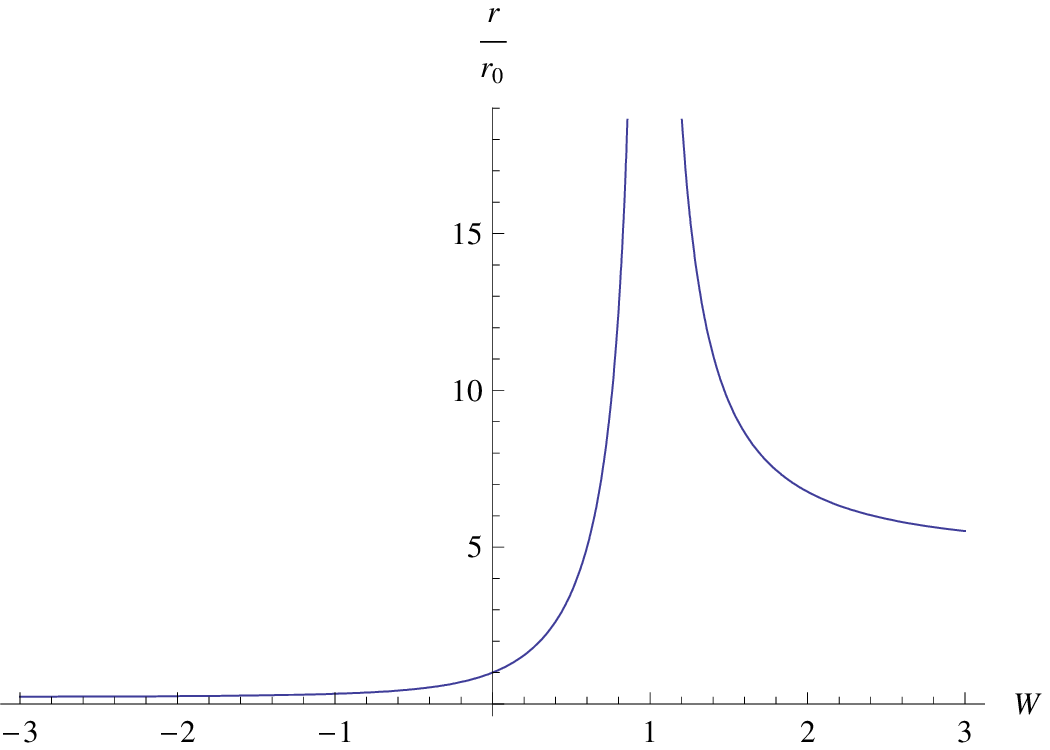}~~~~~\includegraphics[width=7cm]{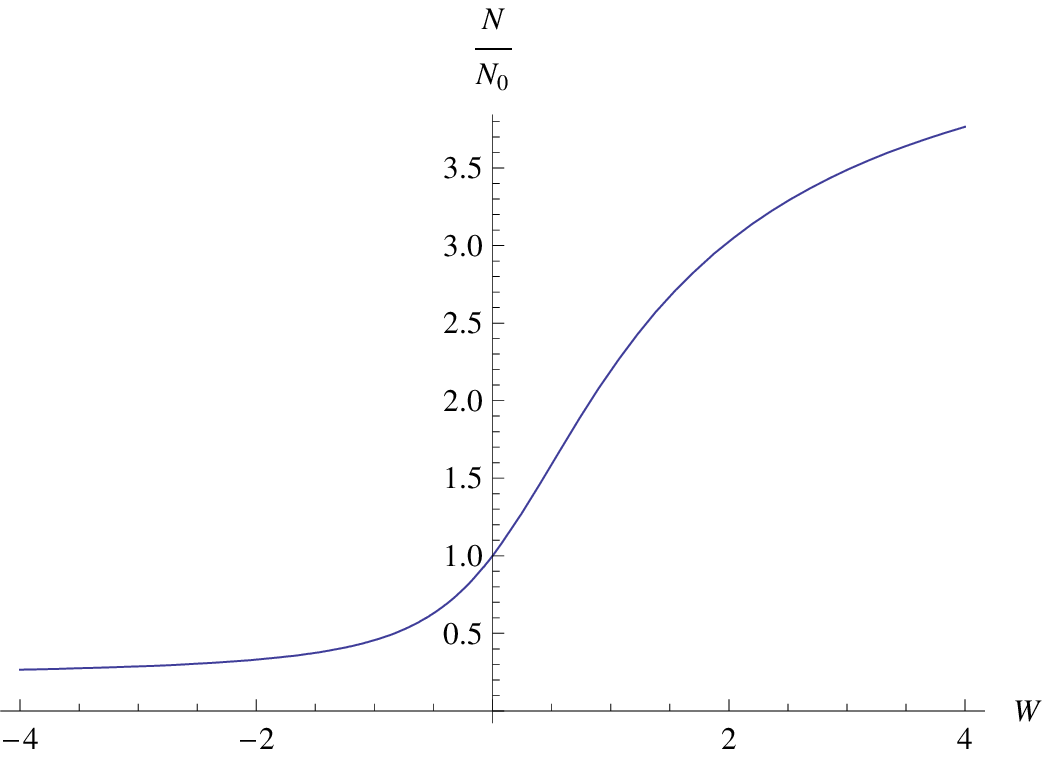}
\end{center}
\caption{\em On the left $r/r_0$  versus $W$ for $a=b=1$. On the right $N/N_0$ versus $W$ for $a=b=1$.}
\label{fig3}
\end{figure}

The solution contains two branches that are visible in figure \ref{fig3}.

Substituting for $g$ we obtain
\begin{itemize}

\item For $W=\sqrt{-ag\over 2+ag}$ For this to be real $0>ag>-2$.

\be
{\sqrt{2\over 2+ag}\exp\left[\sqrt{b\over a}\arctan\left(\sqrt{-ag\over 2+ag}\right)\right]\over
\Big |1-\sqrt{a\over b}\sqrt{-ag\over 2+ag}\Big |}={r\over r_0}\sp N=N_0\exp\left[-\sqrt{b\over a}{\arctan\left(\sqrt{-ag\over 2+ag}\right)
}\right]
\label{p34}\ee

\begin{figure}[ht]
\begin{center}
\includegraphics[width=7cm]{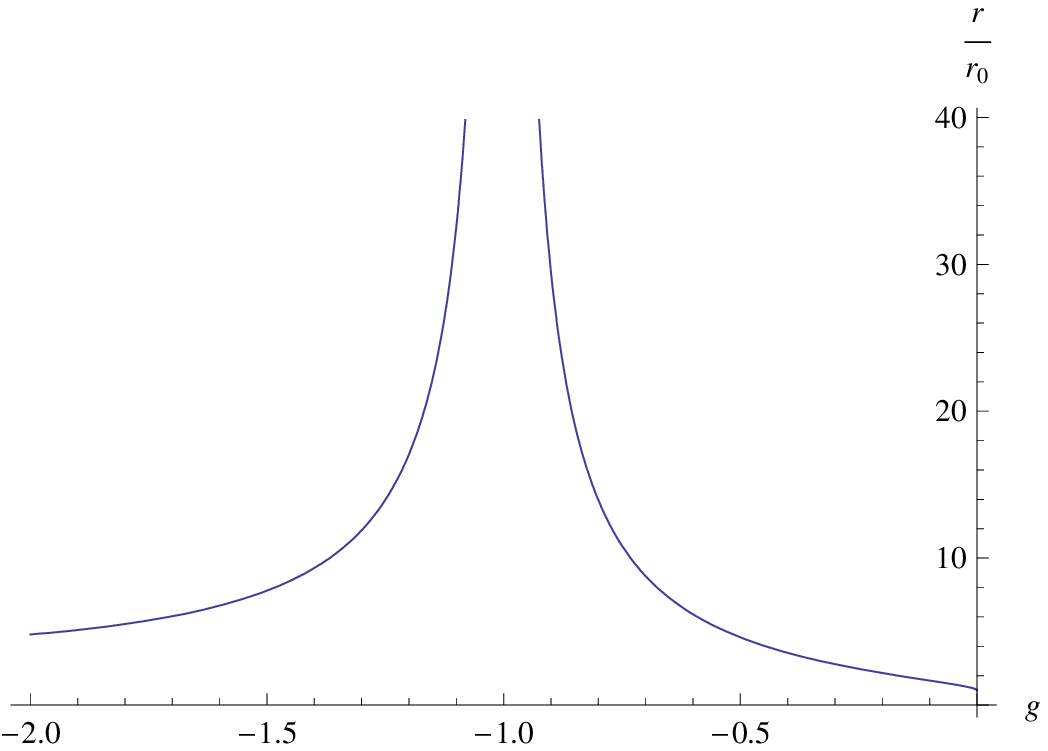}~~~~~
\includegraphics[width=7cm]{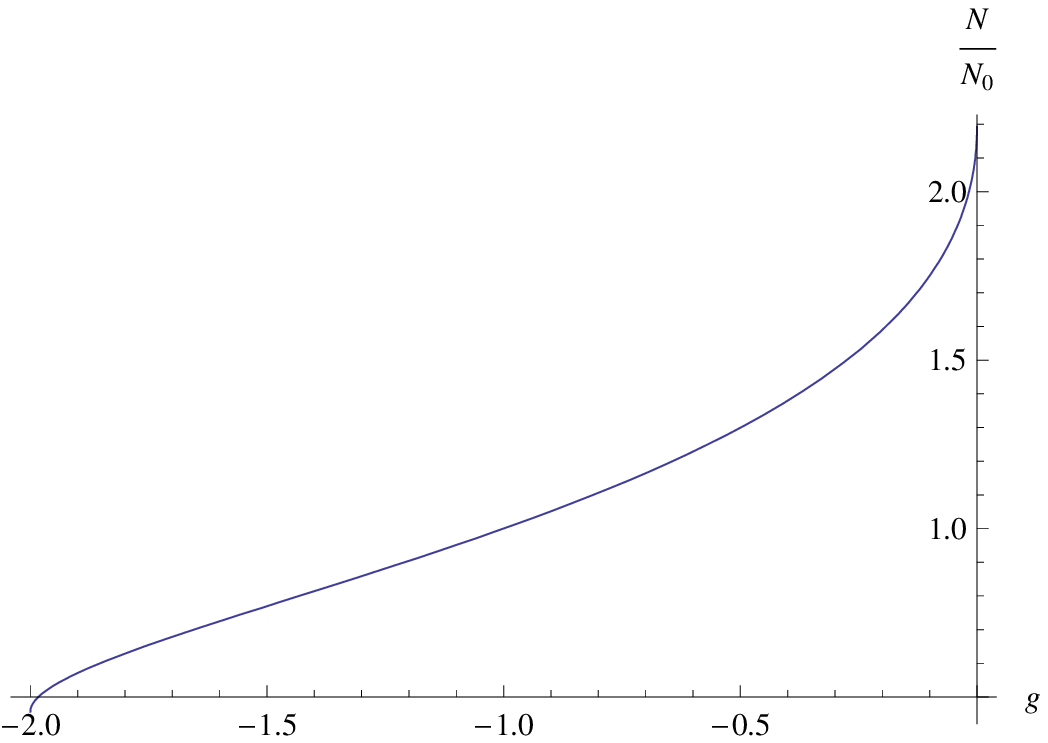}
\end{center}
\caption{\em On the left $r/r_0$ versus $g$, with $W>0$ for $a=b=1$. On the right, N as a function of $g$ normalized to one at $r\to\infty$.}
\label{fig4}
\end{figure}

The functions $g(r)$ and $N(r)$ are shown in figure \ref{fig4}.

\begin{itemize}

\item The left branch $-{2\over a}<g<-{b\over a}$ has an asymptotic infinity at $g=-{b\over a}$ and since it remains smaller than this everywhere
it has a positive mass parameter. It corresponds to $\sqrt{b\over a}<W<\infty$.

\item The right branch $-{b\over a}<g<0$ has an asymptotic infinity at $g=-{b\over a}$ and since it remains larger than this everywhere
it has a negative mass parameter. It corresponds to $0<W<\sqrt{b\over a}$.

\end{itemize}

For the large distance limit $g\to -{b\over a}$ we obtain
\be
g=-{b\over a}+b\sqrt{2\over a}\exp\left[\sqrt{b\over a}\arctan\left(\sqrt{b\over a}\right)\right]{r_0\over r}+{\cal O}(r^{-2})
\label{p35}\ee
\be
N^2=1+b\sqrt{2\over a}\exp\left[\sqrt{b\over a}\arctan\left(\sqrt{b\over a}\right)\right]{r_0\over r}+{\cal O}(r^{-2})
\label{p36}\ee
where we chose $N_0=\exp\left[{\sqrt{b\over a}\arctan\left(\sqrt{b\over a}\right)}\right]$

\item
For $W=-\sqrt{-ag\over 2+ag}$. For this to be real $0>ag>-2$.

\be
{\sqrt{2\over 2+ag}\exp\left[-\sqrt{b\over a}\arctan\left(\sqrt{-ag\over 2+ag}\right)\right]\over
\Big (1+\sqrt{a\over b}\sqrt{-ag\over 2+ag}\Big )}={r\over r_0}\sp N=N_0\exp\left[\sqrt{b\over a}{\arctan\left(\sqrt{-ag\over 2+ag}\right)
}\right]
\label{p344}\ee

\begin{figure}[ht]
\begin{center}
\includegraphics[width=7cm]{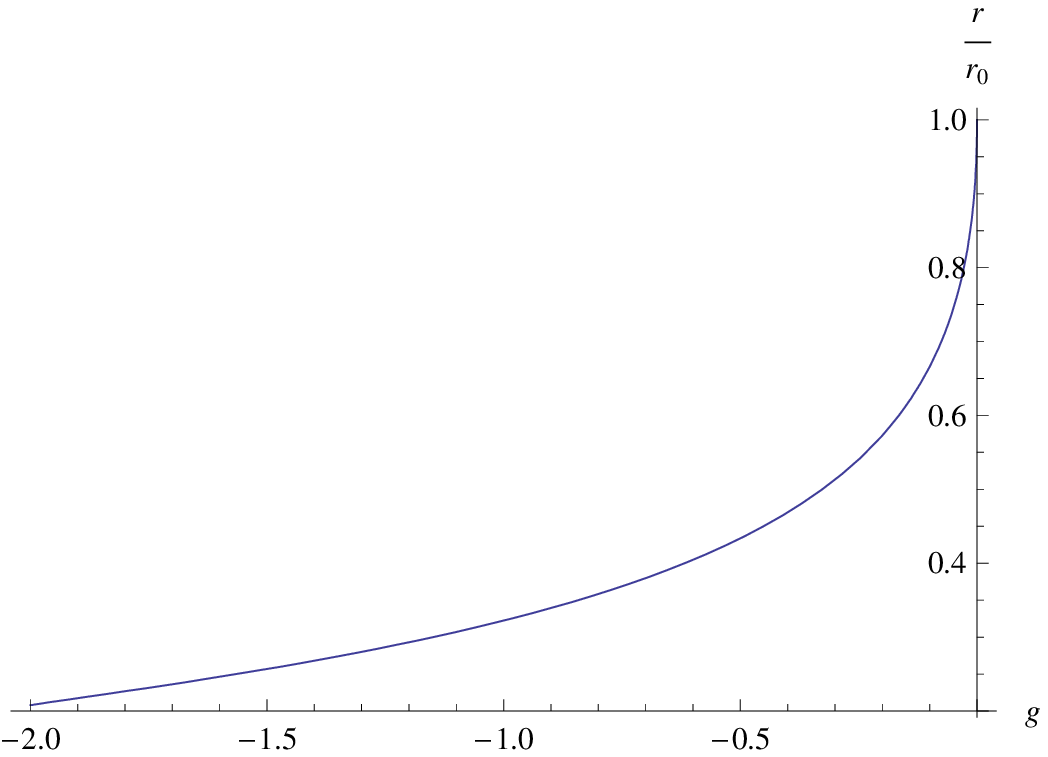}~~~~~
\includegraphics[width=7cm]{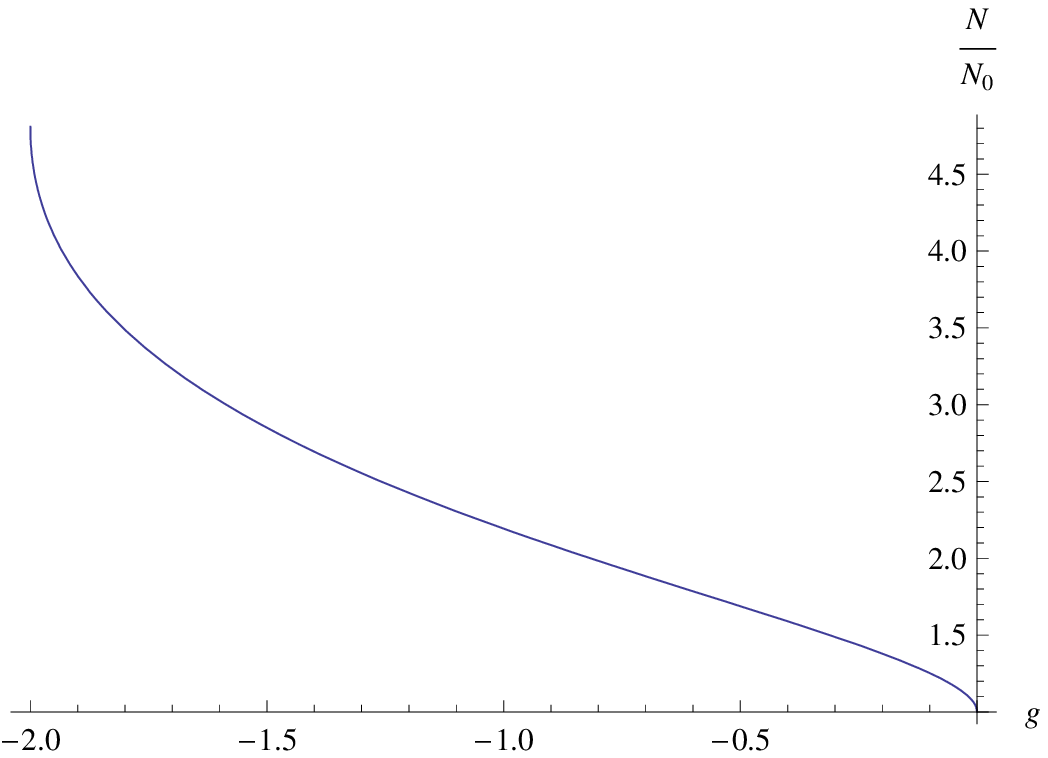}
\end{center}
\caption{\em On the left $r/r_0$ versus $g$, with $W<0$ for $a=b=1$. There is no asymptotic infinity and $r$ varies from 0 to $r_0$.
On the right, $N/N_0$ as a function of $g$.}
\label{fig5}
\end{figure}

Here as shown in figure \ref{fig5} there is no asymptotic infinity. The radius is bounded $0<r<r_0$.
It is quite plausible that this solution combines with the right branch of figure \ref{fig4}, as suggested by figure \ref{fig3}
 in order to make a single solution valid from $0<r<\infty$. This solution has a negative mass parameter.

\end{itemize}


\addcontentsline{toc}{section}{References}
\begin{thebibliography}{99}

\bibitem{hor1}
  P.~Ho\v rava,  {\em ``Quantum Gravity at a Lifshitz Point,''}
    Phys.\ Rev.\  D {\bf 79} (2009) 084008;
  \hri{0901.3775}{[hep-th]};\\


\bibitem{kk}
  E.~Kiritsis and G.~Kofinas,
  {\em ``On Horava-Lifshitz 'Black Holes',''}
  \hri{0910.5487}{[hep-th]}.


  \bibitem{blas2}
  D.~Blas, O.~Pujolas and S.~Sibiryakov,
  {\em ``A healthy extension of Horava gravity,''}
  \hri{0909.3525}{[hep-th]}.


    \bibitem{muko}
  K.~Izumi and S.~Mukohyama,
  {\em ``Stellar center is dynamical in Horava-Lifshitz gravity,''}
  \hri{0911.1814}{[hep-th]}.

 \bibitem{kkco}
  E.~Kiritsis and G.~Kofinas,
  {\em ``Ho\v rava-Lifshitz Cosmology,''}, Nucl. Phys. {\bf B821} (2009) 467-480,
\hri{0904.1334}{[hep-th]}.




\bibitem{papa}
  A.~Papazoglou and T.~P.~Sotiriou,
  {\em ``Strong coupling in extended Horava-Lifshitz gravity,''}
\hri{0911.1299}{[hep-th]}.

\bibitem{Blas3}
  D.~Blas, O.~Pujolas and S.~Sibiryakov,
  {\em ``Comment on `Strong coupling in extended Horava-Lifshitz gravity',''}
  \hri{0912.0550}{[hep-th]}.


\bibitem{E1}
  C.~Eling and T.~Jacobson,
  {\em ``Static post-Newtonian equivalence of GR and gravity with a dynamical
  preferred frame,''}
  Phys.\ Rev.\  D {\bf 69} (2004) 064005
  \hre{gr-qc}{0310044}.

\bibitem{E2}
  C.~Eling and T.~Jacobson,
  {\em ``Spherical Solutions in Einstein-Aether Theory: Static Aether and Stars,''}
  Class.\ Quant.\ Grav.\  {\bf 23} (2006) 5625
  \hre{gr-qc}{0603058}.

\bibitem{E3}
  C.~Eling and T.~Jacobson,
  {\em ``Black holes in Einstein-aether theory,''}
  Class.\ Quant.\ Grav.\  {\bf 23} (2006) 5643
  \hre{gr-qc}{0604088};\\
  T.~Jacobson,
  {\em ``Einstein-aether gravity: a status report,''}
  PoS {\bf QG-PH} (2007) 020
  \hri{0801.1547}{[gr-qc]}.

  \bibitem{Li}
  M.~Li and Y.~Pang,
  {\em ``A Trouble with Ho\v{r}ava-Lifshitz Gravity,''}
  JHEP {\bf 0908} (2009) 015
  \hri{0905.2751}{[hep-th]}.


\bibitem{Hen}
  M.~Henneaux, A.~Kleinschmidt and G.~L.~Gomez,
  {\em ``A dynamical inconsistency of Horava gravity,''}
  \hri{0912.0399}{[hep-th]}.

\bibitem{poly}
  D.~Capasso and A.~P.~Polychronakos,
  {\em ``General static spherically symmetric solutions in Horava gravity,''}
  \hri{0911.1535}{[hep-th]}.


\end{thebibliography}
\end{document}